\documentstyle[epsf,a4,12pt]{article}

\input tcilatex

\begin{document}

\setlength{\baselineskip}{0.8cm}

\begin{titlepage}

\title{The generalised proximity effect model in superconducting
  bi- and trilayer films}

\author{G.~Brammertz$^1$, A.~Poelaert$^1$, A.~Golubov$^2$, P.~Verhoeve$^1$, \\
  A.~Peacock$^1$, H.~Rogalla$^2$. \vspace*{0.3cm} \\
{$^1$\em{Astrophysics Division, Space Science Department,}} \\
{\em European Space Agency, ESTEC,} \\
{\em Noordwijk, The Netherlands} \vspace*{0.3cm}\\
{$^2$\em{Faculty of Applied Physics, Low Temperature Group,}} \\
{\em University of Twente,} \\
{\em Enschede, The Netherlands}}

\date{\today}

\maketitle

\begin{abstract}

\setlength{\baselineskip}{0.8cm}

\setcounter{page}{1}

This paper presents a general model for calculating the density of
states and the Cooper pair potential in proximised superconducting
bi- and trilayer films. It is valid for any kind of bilayer $S_1$-$S_2$, whatever the
quality of the materials $S_1$ and $S_2$, the quality of the
$S_1$-$S_2$ interface and the layer thicknesses. The trilayer model
is valid for a thin $S_3$ layer, whereas the  other two layers have arbitrary
thicknesses. Although the equations of the dirty limit are used, it is argued that the model
stays valid in clean bi- and trilayer films. The typical example of
superconducting tunnel junctions is used to show that existing models,
applying to very thin or very thick layers, or to perfectly
transparent $S_1$-$S_2$ interfaces, are too restrictive to apply to any bilayer.
The new model is applied to existing junctions, with layer
thicknesses intermediate between the 'thick' and the 'thin' approximation.

\end{abstract}

\end{titlepage}

\setcounter{page}{2}

\section{INTRODUCTION}

\label{intro}

Understanding the proximity effect in superconducting films is important for
the development of practical devices such as superconducting tunnel
junctions (STJ's). Depositing a superconductor $S_{1}$ onto another $S_{2}$
modifies the properties of both $S_{1}$ and $S_{2}$ materials. If both
superconductors are thick enough (typically thicker than $10\,\xi
_{S_{1}(S_{2})}$, with $\xi _{S_{1}(S_{2})}$ the coherence length of $S_{1}$
($S_{2}$)), the extremities of the bilayer behave as bulk materials obeying
the BCS theory, though not necessarily like bulk $S_{1}$ and bulk $S_{2}$.
The intermediate region around the $S_{1}$-$S_{2}$ interface is
characterised by a relatively sharp transition between the two bulk-like
regions, and can be pretty far from a BCS-like description. If the layers
are relatively thin, any BCS-like behaviour can be absent from the
structure. Finally, in the case where the layers are extremely thin, as
described by McMillan \cite{mcm}, each layer behaves again like a BCS
superconductor.

The physical quantities affected in a proximised bulk superconductor, are
the Cooper pair potential $\Delta $, the density of states for the Cooper
pairs, $P$, and the density of states for the quasiparticles, $N$. As the
density of states in both superconductors is modified due to the proximity
effect, the resultant bandgap $\Delta _{g}$ lies at an intermediate value
between the bulk values for $S_{1}$ and $S_{2}$, $\Delta _{g,S_{1}}$ and $%
\Delta _{g,S_{2}}$ respectively. This feature has been fully described in
the specific case, of a thin, low bandgap material $S_{2}$ next to a thick,
high bandgap material $S_{1}$, with both superconductors in the dirty limit 
\cite{gol-94,GH-95}.

The goal of the present paper is to present the need for, and develop a
model of, the proximity effect, which is not restricted to this very
specific case.

In particular section~\ref{numeric:applic} shows that there are many
situations where this special case does not apply, and for which the simple
BCS approach does not provide a satisfactory description. Specifically, in
the case of STJ's used as photon detectors, a more general description of
the proximity effect is required to adequately address such issues as device
performance.

The general conditions for an extended model are presented in Section~\ref
{sufficient}. Such a model can be applied not only to STJ's, but also to any
application involving bi- or trilayers of superconductors. The single factor
which would make the model inapplicable is the roughness of the $S_{1}$-$%
S_{2}$ interface. The limitations caused by this parameter are therefore
also discussed.

The extended model is described in Section~\ref{prox}, while numerical
calculations based on the model are presented in Section~\ref{numerical}.
The model is applied to existing STJ's in Section~\ref{numeric:applic}, so
as to highlight a typical application of the model.

\section{The generalised proximity effect model}

\label{prox}

The proximity effect in dirty $S_{1}$-$S_{2}$ bilayers was studied
previously in \cite{GH-95}, \cite{KL}-\cite{Bruder} for the case of
arbitrary transparency of the $S_{1}$-$S_{2}$ interface, but only the case
of a thin $S_{2}$ layer and bulk $S_{1}$ layer was considered. More
recently, the case of arbitrary thickness of $S_{2}$ and $S_{1}$ layers was
considered in \cite{Bruder}, but only the limiting case of the high
transparency of the $S_{1}$-$S_{2}$ interface with $T_{c,S_{2}}=0$ was
studied. Here we generalise the theory to the case of arbitrary $S_{2}$ and $%
S_{1}$ layer thicknesses for the most general case of a finite critical
temperature for the $S_{2}$ layer with $T_{c,S_{2}}<T_{c,S_{1}}$. We also
consider the case of trilayer systems $S_{1}-S_{2}-S_{3}$ with a very thin $%
S_{3}$ layer. The trilayer model remains valid for any thickness of $S_{1}$
and $S_{2}$. The model is constructed for the case of $%
T_{c,S_{3}}<T_{c,S_{1}}$, but $T_{c,S_{2}}$ can have any value.

In case where the relevant length scales (coherence lengths, mean free paths
and sample thicknesses) are much larger than the atomic scale, the proximity
effect can be described in terms of a quasiclassical Green's function
formalism \cite{AGD}-\cite{LO}.

For an arbitrary relationship between the electronic mean free path and the
coherence length the density of states can be derived from the solution of
the Eilenberger equations \cite{Eilenb}. If the vector potential is zero the
Eilenberger equations have the form 
\begin{equation}
{\bf v}_{F}\frac{\partial }{\partial {\bf r}}\widehat{g}(\omega _{n},{\bf r}%
)+\left[ \omega _{n}\widehat{\sigma }_{3}-\Delta ({\bf r})\widehat{\sigma }%
_{1}-\frac{1}{\tau }\left\langle \widehat{g}(\omega _{n},{\bf r}%
)\right\rangle ,\widehat{g}(\omega _{n},{\bf r})\right] =0.  \label{Eilenb}
\end{equation}

In this expression, ${\bf v}_{F}$ is the Fermi velocity, $\widehat{g}=%
\widehat{\sigma }_{3}g+\widehat{\sigma }_{1}f$ is the matrix Green's
function, $\Delta $ is the Cooper pair potential, $\tau $ is the scattering
time of the quasiparticles, and $\widehat{\sigma }_{i}$ are Pauli matrices.
The energy is quantified by the Matsubara frequency, given by $\omega
_{n}=\pi T(2n+1)$ ($T$ is the temperature); the available energies $\epsilon 
$ for the quasiparticles are related to the Matsubara frequencies by the
relation $\omega _{n}=-i\epsilon $. The brackets $\left\langle
...\right\rangle $ denote an averaging over the Fermi surface. In the dirty
limit, the scattering centres ensure that the Green's functions are
isotropic all through the sample. This fact allows Eq.~(\ref{Eilenb}) to be
simplified into a diffusion equation for a Green's function $\widehat{G}%
(\omega _{n},r)$, which is independent of the angles of the vector ${\bf v}%
_{F}$. In this case the function $\widehat{g}$ may be represented by the
form \cite{LO} 
\begin{equation}
\widehat{g}(\omega _{n},{\bf r})=\widehat{G}(\omega _{n},r)+{\bf n}\widehat{G%
}_{1}(\omega _{n},r)\textstyle{,}\;\;\widehat{G}_{1}\ll \widehat{G}\textstyle%
{,}  \label{dirty}
\end{equation}
${\bf n}$ being the unit vector in the direction of the momentum of the
quasiparticles.

It can be shown that $\widehat{G}_{1}={\bf v}_{F}\tau \widehat{G}\frac{%
\partial }{\partial {\bf r}}\widehat{G}$ \cite{Usadel,LO}. In the case of a
short mean free path, the quantity ${\bf v}_{F}\tau $ is small, and the
condition $\widehat{G}_{1}\ll \widehat{G}$ which is required in Equation~(%
\ref{dirty}) is satisfied. This is the condition which is realised either in
the case of a high concentration of impurities in a bulk superconductor or
dislocation scattering centres, or in the case of strong diffusive boundary
scattering in a film.

We consider a superconducting film $S_{i}$ in which the dirty limit
condition $l_{S_{i}}\leq \xi _{S_{i}}$ is fulfilled. Here $l_{S_{i}}$ is the
electronic mean free path. The coherence length $\xi _{S_{i}}$ is related to
the diffusion coefficient $D_{S_{i}}$ by the relation $\xi _{S_{i}}=\sqrt{%
D_{S_{i}}/2\pi T_{c,S_{i}}}$. We shall define the x axis as perpendicular to
the film surface. The Usadel equations \cite{Usadel}, directly derived from
Eq.\ref{Eilenb} can be written as 
\begin{equation}
\xi _{S_{i}}^{2}\theta _{S_{i}}^{^{\prime \prime }}(x)+i\epsilon \sin \theta
_{S_{i}}(x)+\Delta _{S_{i}}(x)\cos \theta _{S_{i}}(x)=0  \label{Us}
\end{equation}
where the pair potential $\Delta _{S_{i}}$ is determined by the
self-consistency relation 
\begin{equation}
\Delta _{S_{i}}(x)\ln \frac{T}{T_{c,S_{i}}}+2T\sum_{\omega _{n}}\left[ \frac{%
\Delta _{S_{i}}(x)}{\omega _{n}}-\sin \theta _{S_{i}}(i\omega _{n},x)\right]
=0.  \label{self}
\end{equation}
Here the function $\theta _{S_{i}}$ has been introduced as a unique Green's
function defining the quasiparticle density of states $N_{S_{i}}$: 
\begin{equation}
N_{S_{i}}(\epsilon ,x)/N_{S_{i}}(0)=Re\left[ \cos \theta _{S_{i}}(\epsilon
,x)\right]  \label{N}
\end{equation}

where $N_{S_{i}}(0)$ is the electronic density of states in the normal state
at the Fermi surface.

In a multilayer structure the Usadel equations Eqs.\ref{Us}, \ref{self} must
be solved in each layer with the use of the appropriate boundary conditions.
For convenience we substitute the coherence length $\xi _{S_{i}}$ by the
quantity $\xi _{S_{i}}^{\ast }=\xi _{S_{i}}\sqrt{T_{c,S_{i}}/T_{c,S_{1}}}$,
normalised to the critical temperature of the layer $S_{1}$.

\subsection{Bilayer $S_{1}-S_{2}$}

Let us first consider the case of a bilayer. For any $S_{1}$-$S_{2}$ system
with $T_{c,S_{2}}<T_{c,S_{1}}$, the origin of the coordinates $x=0$ is taken
at the $S_{1}$-$S_{2}$ interface. The region with $x>0$ is $S_{1}$ and $x<0$
corresponds to $S_{2}$. The layers $S_{1}$and $S_{2}$ have a thickness $%
d_{S_{1}}$ and $d_{S_{2}},$ respectively. The boundary conditions can easily
be written in terms of $\theta _{S_{1}(S_{2})}$ \cite{KL}. At the $S_{1}$-$%
S_{2}$ interface we have 
\begin{equation}
\gamma _{BN}\xi _{S_{2}}^{\ast }\theta _{S_{2}}^{\prime }=\sin (\theta
_{S_{1}}-\theta _{S_{2}})  \label{BC1}
\end{equation}
\begin{equation}
\gamma \xi _{S_{2}}^{\ast }\theta _{S_{2}}^{\prime }=\xi _{S_{1}}\theta
_{S_{1}}^{\prime }.  \label{BC2}
\end{equation}

At the free interface of both $S_1$ and $S_2$ layers, the conditions are:

\begin{equation}
\theta _{S_2}^{^{\prime }}(-d_{S_2})=0 \;\; \textstyle{;} \;\; \theta
_{S_1}^{^{\prime }}(d_{S_1})=0.  \label{BNS}
\end{equation}

The parameters $\gamma _{BN}$ and $\gamma $ involved in the boundary
conditions at the $S_{1}$-$S_{2}$ interface are given by 
\begin{equation}
\gamma _{BN}=\frac{R_{B}}{\rho _{S_{2}}\xi _{S_{2}}^{\ast }}\;\;\textstyle{;}%
\;\;\gamma =\frac{\rho _{S_{1}}\xi _{S_{1}}}{\rho _{S_{2}}\xi _{S_{2}}^{\ast
}}.  \label{Gamma}
\end{equation}
These parameters can be understood as follows: $\gamma $ is a measure of the
strength of the proximity effect between the $S_{1}$ and $S_{2}$ metals,
whereas $\gamma _{BN}$ describes the effect of the boundary transparency
between these layers. Here $\rho _{S_{1},S_{2}}$ are normal state
resistivities and $R_{B}$ is the product of the resistance of the $S_{1}$-$%
S_{2}$ boundary and its area.

As is shown in Ref.\cite{GH-95}, in the case of a thin $S_{2}$ layer where $%
d_{S_{2}}/\xi _{S_{2}}^{\ast }\ll 1$, the parameters defining the proximity
effect are $\gamma _{m}=\alpha \gamma d_{S_{2}}/\xi _{S_{2}}^{\ast }$ and $%
\gamma _{B}=\alpha \gamma _{B}d_{S_{2}}/\xi _{S_{2}}^{\ast }$. Here $\alpha $
is a correction factor which is a function of the ratio of critical
temperatures $T_{c,S_{2}}/T_{c,S_{1}}$, and of the thickness of the layer $%
S_{2}$. The dependence of the parameter $\alpha $ on $T_{c,{S_{2}}}/T_{c,{%
S_{1}}}$ can be found in Ref.\cite{GH-95}.

\subsection{Trilayer $S_{1}-S_{2}-S_{3}$}

We consider now the trilayer situation. This case is rather straightforward
to solve with only few conditions. We first assume $d_{S_{3}}\ll \xi
_{S_{3}}^{\ast }$.Again the origin of the coordinates $x=0$ is taken at the $%
S_{1}$-$S_{2}$ interface and the $S_{3}$ layer occupies the region $%
-d_{S_{2}}-d_{S_{3}}<x<-$ $d_{S_{2}}$. As mentioned in \cite{gol-94}, this
approximation directly implies that pair potential and density of states in $%
S_{3}$ are constant through its whole thickness. The solution in the thin $%
S_{3}$ film is \cite{gol-94}

\begin{equation}
\tan \theta _{S_{3}}=\frac{\sin \theta _{S_{2}}(-d_{S_{2}})+\gamma
_{BN_{2}}\Delta _{S_{3}}}{\cos \theta _{S_{2}}(-d_{S_{2}})+\gamma
_{BN_{2}}\omega }  \label{tri}
\end{equation}

The boundary conditions at the interface $S_{1}$-$S_{2}$ and at the free
surface of $S_{1}$ can be derived from the general bilayer problem described
above

\begin{equation}
\gamma _{BN_{1}}\xi _{S_{2}}^{\ast }\theta _{S_{2}}^{\prime }=\sin (\theta
_{S_{1}}-\theta _{S_{2}})  \label{BC1}
\end{equation}
\begin{equation}
\gamma _{1}\xi _{S_{2}}^{\ast }\theta _{S_{2}}^{\prime }=\xi _{S_{1}}\theta
_{S_{1}}^{\prime }.  \label{BC2}
\end{equation}

\begin{equation}
\theta _{S_1}^{^{\prime }}(d_{S_1})=0.  \label{BNS}
\end{equation}

An additional boundary condition can be introduced at the $S_{2}$-$S_{3}$
interface, based on the fact that $d_{S_{3}}$ is very small

\begin{equation}
\xi _{S_{2}}^{\ast }\theta _{S_{2}}^{\prime }(-d_{S_{2}})=\frac{\gamma
_{2}\omega \left\{ \sin \theta _{S_{2}}(-d_{S_{2}})-\Delta _{S_{3}}\right\} 
}{\left\{ 1+\gamma _{BN_{2}}^{2}(\omega ^{2}+\Delta _{S_{3}}^{2})+2\gamma
_{BN_{2}}^{{}}\cos \theta _{S_{2}}(-d_{S_{2}})\left[ \omega +\Delta
_{S_{3}}\sin \theta _{S_{2}}(-d_{S_{2}})/\omega \right] \right\} }
\label{bc3}
\end{equation}

Here the parameters $\gamma _{BN_{i}}$ and $\gamma _{i}$ are given by 
\begin{equation}
\gamma _{BN_{i}}=\frac{R_{Bi}}{\rho _{S_{i+1}}\xi _{S_{i+1}}^{\ast }}\;\;%
\textstyle{;}\;\;\gamma _{i}=\frac{\rho _{S_{i}}\xi _{S_{i}}^{\ast }}{\rho
_{S_{i+1}}\xi _{S_{i+1}}^{\ast }}.  \label{Gamma3}
\end{equation}

Again, the coherence lengths in $S_{2}$ and $S_{3}$ are normalized to $%
T_{c,S_{1}}$. Using a similar approach to those adopted previously, it is
also straightforward to study the case of a very thin $S_{1}$ layer with
arbitrary $S_{2}$ and $S_{3}$ film thicknesses.

\section{Applicability of the model to the clean limit}

\label{sufficient}

The main restriction of the model presented here is the use of dirty
superconductors. It is straigthforward to show that, in most cases, the
clean limit does not need to be considered. The difference between clean and
dirty limits lies in the presence of impurities or crystallographic
dislocations in a dirty superconductor, acting as scattering centres for
quasiparticles. If scattering centres are on average separated by a distance
smaller than the coherence length, then the dirty limit applies. In other
words, the dirty limit localy applies to a spherical region of radius equal
to the coherence length, and centered at a scattering centre.

The physical presence of interfaces ensures the presence of scattering
centres. As long as the interfaces affect several atomic layers of both
superconductors, the boundary conditions are those of the dirty limit.
Moreover, over a scale length equal to the coherence length in both films,
any solutions for the pair potential and the densities of states are also
governed by the same equations valid in a dirty superconductor. At depths
larger than the coherence length, the fluctuations in pair potential and in
densities of states are not significant, and the behaviour tends towards
that of a bulk superconductor \cite{gol-94,GH-95}. Anderson's theorem \cite
{anderson} implies that in the case of a bulk superconductor, the solutions
in the clean limit are the same as in the dirty limit. Hence, any solution
for the pair potential and the density of states in regions of the films
deeper than the coherence length, is expected to be a smooth interpolation
between two dirty regions (in the case of a clean region separated by two
rough boundaries), or a smooth extrapolation out of a single dirty region
(in the case of a clean region located between a rough boundary and a flat
one). The solutions related to such smooth interpolations or extrapolations
are identical, independently of the use of the equations valid for the clean
or for the dirty limit.

The clean limit only need to be considered in cases of atomically sharp
inferfaces, where the roughness $\sigma _{i}$ of the $S_{1}$-$S_{2}$
interface is of the order of or smaller than the interatomic distance of
both $S_{1}$ and $S_{2}$ (typically around 4-5~\AA ). This case is extremely
restrictive, and also technically difficult to achieve, even with the most
advanced methods of epitaxial thin film depositions. It leads to the
conclusion that the clean limit does not need to be considered.

\section{Numerical results}

\label{numerical}

In this section some typical examples of pair potential and density of
states for the quasiparticles are determined and discussed for bi- and
trilayers.\bigskip

\subsection{Bilayer $S_{1}-S_{2}$}

\label{bilayer}For this bilayer, we have solved numerically the Usadel
equations (\ref{Us})-(\ref{self}) by a selfconsistent procedure similar to
that described in Ref. \cite{GH-95}, but without the approximation $%
d_{S_{2}}/\xi _{S_{2}}^{\ast }\ll 1$. We start from the trial pair
potentials $\Delta _{S_{1}(S_{2})}(x)$ and find the solutions for $\theta
_{S_{1}(S_{2})}(\omega _{n},x)$ in the Matsubara representation ($\epsilon
=i\omega _{n}$). Using these solutions, the new pair potentials $\Delta
_{S_{1}(S_{2})}(x)$ are found from the selfconsistency equation. The
iterations are repeated until convergence is achieved. Next, after the pair
potentials are determined, we solve the equations (\ref{Us})-(\ref{self}) on
the real energy axis $\epsilon $. According to Equ.(\ref{N}), this method
provides spatially and energy resolved densities of states in both layers.

In order to demonstrate the validity of the approach and to be
representative of the devices discussed in Section \ref{numeric:applic}, the
cases of an Al-Nb and an Al-Ta bilayer ($T_{c,Al}=1.18$~K, $T_{c,Ta}=4.5$ K, 
$T_{c,Nb}=9.25$~K) at $T=0.3$~K are considered.

\subsubsection{$Al-Nb$ bilayer}

\label{Al-Nb}

The parameters for this simulation are $\gamma =1.3$, $\gamma _{BN}=2.7$, $%
d_{Al}=1.7\xi _{Al}^{\ast }$ and $d_{Nb}=4.35$ $\xi _{Nb}$. This choice of
parameters is justified in Section \ref{numeric:applic}. The results are
represented in Fig.1. The quantities represented here are the pair potential
as a function of position in the bilayer (Fig. 1a) and the density of states
at different positions (Fig. 1b). The density of states of Fig. 1b is shown
at the free interface of $S_{1}$ (Nb) and of $S_{2}$ (Al) (solid lines), and
on both sides of the Al-Nb interface (dashed lines). In Al, the density of
states is peaking at an energy slightly higher than the gap, and then
roughly decreases to 1 for infinite energies. Getting closer to the Nb film,
a smaller peak is also visible around the energy gap of pure Nb. In Nb, the
density of states is strongly peaking around $\Delta _{Nb}$ , and goes to 1
at infinite energy. Below $\Delta _{Nb}$the density of states is very much
depressed as compared to the density of states in Al. However, it stays
finite and importantly the energy gap is the same as in the Al film.

The position dependence of the energy gap is represented on Fig. 1a (dashed
line in the middle). Clearly, it is not position dependent at all, despite
the very strong fluctuations of the pair potential. It must be stressed
that, even if Fig. 1b suggests a small density of states in Nb below $\Delta
_{Nb}$, this level actually corresponds to a significant number of states.
As a matter of fact, the density of states at the free interface of Nb
(solid line in Fig. 1b) at 0.3$\Delta _{Nb}$ is $\simeq 1.43\cdot 10^{6}$%
states/meV/${\rm \mu }$m$^{3}$. Despite the fact that the gap is constant,
the reduction in the DOS still has an effect on quasiparticle dynamics,
since the quasiparticles in the energy range $0.3\Delta _{Nb}<\varepsilon
<\Delta _{Nb}$ entering Nb are partially Andreev reflected. As shown in Ref. 
\cite{Aminov} the probability to enter the reduced density of states region
is proportional to $N(\varepsilon )$.

\subsubsection{$Al-Ta$ bilayer}

\label{Al-Ta}

For this simulation the parameters are $\gamma =0.05$, $\gamma _{BN}=3.$, $%
d_{Al}=\xi _{Al}^{\ast }$ and $d_{Ta}=1.18$ $\xi _{Ta}.$ Again the choice of
this set of parameters is explained in Section \ref{numeric:applic}. The
results are represented in Fig 2. The value used for $\gamma _{BN}$ is very
similar to that used for the Nb-based sample, whereas the value used for $%
\gamma $ is about 20 times lower. This has a direct impact on the shape of
the pair potential (Fig. 2a) and the density of states (Fig. 2b). The
discontinuity at the Ta/Al interface is extremely strong. The behaviour of
Ta is almost bulk-like, with a constant pair potential and a steep peak of
the density of states at around $\Delta _{0}^{Ta}$. However there is still a
considerable amount of states available in Ta below $\Delta _{0}^{Ta}$
(typically 0.1 $\times $ 2$N_{0}$ =8.1 10$^{6}$ states/meV/${\rm \mu m}^{3}$%
). The fluctuations of the pair potential and the density of states within
Al are also rather small, though a clear contribution from Ta remains
visible at $\Delta _{0}^{Ta}$. Despite the very weak coupling between Al and
Ta, the gap is still constant through the whole device thickness. Such a
strong discontinuity at the interface emphasizes the role of quasiparticle
confinement in Al, due to Andreev reflections.

\subsection{Trilayer $S_{1}-S_{2}-S_{3}$}

Next we consider two representative examples of a trilayer system: Al-Nb-NbN
and Al-Nb-Ta, which also do have practical applications. The inclusion of a
thin NbN passivation layer on the top electrode of an Al-Nb STJ in place of
the natural niobium oxides reduces the quasiparticle loss rate, thereby
enhancing the probability of multiple tunnel processes \cite{NbN}. On the
other hand the deposition of a thin Nb layer on top of the Al facilitates
the deposition of the top Ta in Al-Ta STJs. The best Ta based devices, in
terms of detection of optical photons, have a top electrode of this kind.

In both cases the Al layer has a thickness $d_{Al}=0.1\xi _{Al}^{\ast }$ ,
for which the thin layer approximation works reasonably well, while no
limitation on the thickness of other layers was introduced.The method of
numerical solution in this case is similar to that for a bilayer with
additional boundary conditions (\ref{bc3}), as described in the previous
section.

\subsubsection{$Al-Nb-NbN$ trilayer}

In the case of Al-Nb-NbN we have taken thicknesses $d_{Nb}/\xi _{Nb}^{\ast
}=3.$ and $d_{NbN}/\xi _{NbN}^{{}}=3$, and $T_{c,NbN}=14K$. The parameters
of the Al-Nb interface are $\gamma _{2}=1.3$ and $\gamma _{BN_{2}}=2.7$, as
in the previous example. For the Nb-NbN interface $\gamma _{1}=2$ is
expected from the resistivity ratio, while for the barrier transparency
parameter we have assumed $\gamma _{BN_{1}}=1$. The results of calculations
of the spatial distribution of the pair potential in the Al-Nb-NbN trilayer
are shown in Fig.3a. The dashed line shows the energy gap, which is again
constant across the trilayer. The densities of states are shown in Fig.3b at
the free interface of NbN, $x=3\xi _{NbN},$ (solid line), on both sides of
the NbN-Nb interface, $x=\pm 0,$ (dashed lines), on the Nb side of the Nb-Al
interface, $x=-3\xi _{Nb}$ (solid line) and in Al (dotted line). Again, in
Al the density of states is peaking at an energy slightly higher than the
gap. The magnitude of the gap is higher than in the case of Fig.1 due to the
smaller thickness of Al and the influence of the NbN layer. Note, the gap in
Al is now as big as about half of the gap of pure NbN, $\Delta _{NbN}$. The
density of states at the free interface of NbN, $x=3\xi _{NbN},$ peaks
around $\Delta _{NbN}$, while a lot of states are also present in this layer
at $\varepsilon <\Delta _{NbN}.$

\subsubsection{$Al-Nb-Ta$ trilayer}

In the case of Al-Nb-Ta we have taken thicknesses $d_{Nb}/\xi _{Nb}^{\ast
}=0.5$ and $d_{Ta}/\xi _{Ta}^{{}}=3$, and $T_{c,Ta}=4.47K$. The parameters
of the Nb-Ta interface are assumed to be $\gamma _{1}=0.3$ and $\gamma
_{BN_{1}}=3.8$. For the Al-Nb interface $\gamma _{2}=1.3$ and $\gamma
_{BN_{2}}=2.7$ were chosen again. The distribution of the pair potential,
the energy gap and the local densities of states are shown in Fig.4a,b. The
notations in Fig.4b are the same as in Fig.3b with NbN and Ta interchanged.
As is seen from comparison of Figs.3 and 4, the energy gap in the Al-Nb-Ta
trilayer is lower than that in the Al-Nb-NbN. This is expected, since the
only difference between these two cases is that the high gap material, NbN,
is substituted by the low gap one, Ta. Moreover, the peaks of the density of
states in Nb are sharper in the case of Al-Nb-Ta trilayer. The reason is the
more homogeneous distribution of the pair potential in this case, as is seen
from Fig.4a.

\section{Application}

\label{numeric:applic}

The STJ would represent a typical but not exclusive example of an
application of multi-layered superconducting thin films. Such devices used
as photon detectors have produced good results in terms of energy
resolution, charge output and quantum efficiency, for photon energies from
the near infrared to X-rays \cite{rando92}-\cite{ver97}. The most important
results currently achieved can be summarised as follows: Ta/Al junctions
provide a typical response of up to {$10^{5}$} electrons per eV of the
detected photon, at a temperature of 300 to 400~mK; the associated measured
energy resolution is within a few percents of the predicted theoretical
limit; the quantum efficiency at UV wavelengths is about 60\%; the longest
wavelength detected is currently 2{${\rm \mu m}$}, with the possibility to
extend this limit to 10{${\rm \mu m}$}; the same junctions operate also at
x-ray energies as high as 6 keV, with a quantum efficiency of about 10\% and
an energy resolution of about 40 eV FWHM. Very encouraging results have been
also achieved with Nb/Al \cite{ver97} and NbN/Nb \cite{NbN} STJ's. At x-ray
energies (6 keV), the best results currently achieved yield an energy
resolution of 12~eV for 5.9~keV photons, with an Al-based junction working
at 50~mK \cite{hettl}.

A major problem to be addressed is the ability to predict the junctions
response to photon absorption. This response can be modified by changing
such variables as film thickness, film quality, and overall device
dimensions. Especially, the performance of existing devices is still
difficult to predict when the proximity effect plays a significant role. In
particular, a significant difference in junction behaviour has been observed
when irradiated by x-rays, as opposed to optical photons (\cite{ver97}).
Note while the trend in X-ray responsivity reflects the generally increasing
role of quasiparticle self-recombination with increasing photon energies,
the details of this response have been difficult to model using existing
theories. Also, adressing adequately the problems of energy non-linearity (
\cite{goldie}-\cite{kaplan}) and quasiparticle trapping (\cite{booth})
requires a detailed knowledge of the proximity effect theory.

\subsection{Applicability of previous models}

The aim of this section is to show that the existing models for the
proximity effect are for very specific cases and regimes only and need
generalisation in order to explain the performance of an STJ.

The type of junctions in which the energy dependence of the responsivity is
difficult to account for are described in \cite{poelaert96}. They
essentially consist of symmetrical Nb/Al/AlO$_{x}$/Al/Nb junctions deposited
on super-polished $R$-plane sapphire, with 100~nm of Nb and 15 to 120~nm of
Al. The base electrode is epitaxial, whilst the top is polycrystalline. The
polycrystalline film is usually covered by a thin NbO layer, followed by a Nb%
$_{2}$O$_{5}$ layer ($\sim 5$~nm in total), due to the normal exposure to
air. Note, the bandgap is much smaller in Al than in Nb. The Al layer
represents the $S_{2}$ layer, and the Nb the $S_{1}$ layer.

A typical value for the coherence length in bulk, clean Al at 0~K is $\xi
_{Al}(0)\sim $~1.6~${\rm \mu m}$. For the polycrystalline film (superscripts 
$p$ indicates polycrystalline), the electron mean free path $l_{Al}^{p}$ is
limited by the grain size. Using a Transmission Electron Microscope (TEM),
the grain size in the polycrystalline Al film has been estimated at about
40~nm. In the dirty limit, the coherence length is given by $\xi _{Al}^{p}=%
\sqrt{l_{Al}^{p}\xi _{Al}(0)/3}\simeq $~147~nm, and $\xi _{Al}^{\ast ,p}=\xi
_{Al}^{p}\sqrt{T_{c,Al}/T_{c,Nb}}\simeq $~53~nm. Thus the condition of the
dirty limit for the polycrystalline films $S_{1}$ and $S_{2}$ is not
actually valid. Furthermore, this coherence length is precisely in the
middle of the range of Al thicknesses available, and the condition of small
Al thickness is also not satisfied. As for the epitaxial film (superscripts $%
e$), the mean free path $l_{Al}^{e}$ is not limited by the grain size
anymore, but it is simply constrained by the boundary scattering due to the
thickness of the electrode (220~nm) and thus is always far lower than $\xi
_{Al}(0)$. This conclusion is consistent with $RRR$ (resistance residual
ratio) measurements, which lead to an average mean free path of $\sim $%
~150~nm over the whole electrode. The coherence length $\xi _{Al}^{\ast ,e}$
can therefore be assumed to be governed by a dirty environment, and will not
exceed $\xi _{Al}^{\ast ,p}\sqrt{l_{Al}^{e}/l_{Al}^{p}}<120$~nm. This is not
very large compared to either $\l _{Al}^{e}$ or $d_{Al}$. As a consequence,
in the epitaxial film, none of the assumptions associated with the dirty
limit and with a thin Al layer are valid.

In Nb, $\xi _{Nb}(0)\sim $~38~nm. This is very close to the grain size in
the polycrystalline film, implying that the Nb polycrystalline film is in
between the dirty and the clean limit. If the layer $S_{1}$ was assumed to
be dirty, the coherence length is $\xi _{Nb}^{p}\sim $~23~nm. In reality, it
is somewhere between 23 and 38~nm. In any case, the dirty limit does not
apply. In the epitaxial layer, $RRR$ measurements have determined an
electron mean free path $l_{Nb}^{e}\simeq $~150~nm; here we are clearly in
the clean limit, constrained by boundary scattering, such that for a Nb film
of thickness 100~nm (applicable for all examples described herein), the
condition of a thick $S_{1}$ layer is not fulfilled.

\subsection{Application of the generalised proximity effect model}

The model has been applied to two different STJ layups. The first one is a
symetrical Nb based STJ with 100nm of Nb and 90 nm of Al. The second is a
symetrical Ta based STJ with 100nm of Ta and 55 nm of Al. In order to
determine the density of states and the order parameters, all the input
parameters for the proximity effect model have to be established. These
parameters are the critical temperature, the ratio between the thickness of
each layer and the coherence length, and the interface parameters.

The coherence length in a non bulk configuration can be expressed in terms
of the bulk coherence length $\xi _{0}$ and of the electron mean free path $%
l $. If the superconductor is in the clean limit, this relation is

\begin{equation}
\frac{1}{\xi }=\frac{1}{\xi _{0}}+\frac{1}{l}\text{.}  \label{colecl}
\end{equation}

In the dirty limit, where $l$ \TEXTsymbol{<}\TEXTsymbol{<} $\xi $, one has

\begin{equation}
\xi =\sqrt{\frac{hD}{2\pi kT_{c}}}\text{.}  \label{coledi}
\end{equation}

Assuming the clean limit in epitaxial base films and the dirty limit in
polycrystalline top films, limiting the mean free path to the grain size (40
nm) in polycrystalline films, the values of $\xi _{Nb}$ \symbol{126}23 nm, $%
\xi _{Al}$ \symbol{126}147 nm and $\xi _{Ta}$ \symbol{126}85 nm were found.
These values are actually averages between the values found for
polycrystalline and epitaxial films.Using these values we determine the
ratios of thickness to normalised coherence length as used for the numerical
simulations in Sections \ref{numerical} \ref{bilayer} \ref{Al-Nb} and \ref
{Al-Ta}.

The method chosen to determine the values of the interface parameters is to
compare an experimentally determined value with the theoretical estimate, as
derived from a simple analytical expression for this value, which includes
the densities of states. The quantities to be fitted to the model were
chosen to be the energy gap as a function of temperature and the critical
current as a function of temperature. Both quantities are easily found
experimentally from the junctions I-V curve characteristics. The critical
current $I_{c}$ through the junction is calculated in the following way\cite
{GH-95} 
\begin{equation}
\frac{eI_{c}R_{N}}{2\pi T_{c}}=\frac{T}{T_{c}}\sum_{\omega _{n}>0}\sin
\theta _{1}(\omega _{n})\sin \theta _{2}(\omega _{n}),  \label{super}
\end{equation}
where $R_{n}$ is the normal-state resistance of the junction, and $\theta
_{1(2)}(\omega _{n})$ is the local Green's function in the vicinity of the
barrier in electrode 1 and 2 respectively. Both electrodes are assumed to be
identical, therefore simplifying Eq.~(\ref{super}), since $F_{1}=F_{2}$, and
thereby ensuring a unique set of $(\gamma ,\gamma _{BN})$ parameters for the
whole junction.

Figs. 5a and 6a show the model values compared to the experimental data of
the energy gap as a function of device temperature for the Nb and the Ta
based STJs respectively. Figs 5b and 6b show essentially the same
information, but this time in regard to the critical current.

The fits to the experimental points correspond to model values with $\gamma
=1.3$, $\gamma _{BN}=2.7$ for the Nb based STJ and $\gamma =0.05$, $\gamma
_{BN}=3$ for the Ta based STJ. In both cases the fit of the bandgap is
satisfactory, while the critical current of the Nb based device is slightly
too low.

Using these model derived values of the parameters we obtain the results
presented in sections \ref{numerical} \ref{bilayer} \ref{Al-Nb} and \ref
{Al-Ta}.

\subsection{Dependence of the interface parameters on film thickness}

We will now analyse how the interface parameters vary when the thicknesses
of the films change, all other parameters staying the same. Assuming that
the dirty limit is valid, we can replace the coherence length by its
expression in the dirty limit:

\begin{equation}
\xi =\sqrt{\frac{\xi _{0}l}{3}}  \label{coledi2}
\end{equation}

This gives the following expressions for the interface parameters:

\begin{equation}
\gamma _{BN}=C_{\gamma _{BN}}\sqrt{\frac{T_{c,1}l_{2}}{T_{c,2}}}\;\;%
\textstyle{;}\;\;\gamma =C_{\gamma }\sqrt{\frac{T_{c,1}l_{2}}{T_{c,2}l_{1}}}.
\label{Gammabis}
\end{equation}

where the constants C$_{\gamma }$ and C$_{\gamma _{BN}}$ are defined by:

\begin{equation}
C_{\gamma _{BN}}=\frac{R_{B}}{\rho _{2}l_{2}}\sqrt{\frac{3}{\xi _{0,2}}}\;\;%
\textstyle{;}\;\;C_{\gamma }=\frac{\rho _{1}l_{1}}{\rho _{2}l_{2}}\sqrt{%
\frac{\xi _{0,1}}{\xi _{0,2}}}.  \label{Gamcon}
\end{equation}

These two constants are independent of the film thickness. C$_{\gamma }$
depends only on the nature of the two films involved, whereas C$_{\gamma
_{BN}}$ depends also on the macroscopic properties of the interface. The
only variables depending on the film thickness are the critical temperature
and the mean free path.

The critical temperature as a function of film thickness can be determined
according to the model of Cooper \cite{Cooper}. This model states that
superconductivity is lost in a thin surface layer b$_{t}$ due to a reduction
in the electron density of states near the surface. The critical temperature
dependence according to this model is:

\begin{equation}
T_{c}=T_{c,0}\left( 1-\frac{2b_{t}}{N\nu t}\right) ,  \label{Tc}
\end{equation}

where T$_{c,0}$ is the critical temperature of the bulk material, N is the
electron density of states at the Fermi level, $\nu $ the bulk interaction
potential and t the film thickness.

The bulk mean free path at low temperature l$_{0}$ (just above the critical
temperature) can be found via the bulk mean free path at 300K, l$_{0,300}$,
and the residual resistance ratio (RRR$_{b}$) of a thick film:

\begin{equation}
l_{0}=RRR_{b}l_{0,300}.  \label{RRR}
\end{equation}

We can then calculate the mean free path in a thin film (t \TEXTsymbol{<}%
\TEXTsymbol{<} l$_{0}$), using \cite{Movshovitz}:

\begin{equation}
l=\frac{3t}{4}\left( \ln \frac{l_{0}}{t}+0.423\right) .  \label{mfp_vs_t}
\end{equation}

Using the interface parameters found for a Nb based STJ with 45 nm of Al and
the basic material constants for Nb and Al (table 1) appearing in equations (%
\ref{coledi2})-(\ref{mfp_vs_t}), we can determine the interface constants $%
C_{\gamma }$ and $C_{\gamma _{BN}}.$ Except for the factor N$\nu $ (cfr. eq. 
\ref{Tc}), which was determined in Ref. \cite{Meservey}, all values in table
1 were derived from experiments on Nb thin films or Nb/Al/AlOx/Al/Nb
multilayers. For these films the critical temperature and RRR were measured
as a function of film thicknesses. A fit to these experimental points
provided the values of b$_{t}$ and T$_{c,0}$.

Using equations (\ref{Tc}) and (\ref{mfp_vs_t}) we then calculate the values
of the interface parameters for different Al film thicknesses. The
theoretical and experimental values of the interface parameters for
symmetrical Nb/Al STJs with Al film thicknesses of 15, 30, 45 and 90 nm and
a constant Nb film thickness of 100nm can be found in table 2. The
experimental values were determined in exactly the same way as described in
the previous section. The correspondence between experiment and theory is
very good, when we take into account the rather large uncertainties we are
dealing with respect to the numerous parameters involved. Note that
including the $\gamma _{BN}$ parameter in the discussion is important, as it
is not small and has a strong effect on the density of states in the two
films. This is in accordance with the results obtained by Zehnder et al. (
\cite{zehnder}). Nevertheless the dependence on film thickness is rather
complex for both interface parameters. No clear square root dependence for $%
\gamma $ or linear dependence for $\gamma _{BN}$ was observed, as stated in
ref \cite{zehnder}.

\section{CONCLUSIONS}

\label{ccl}

A model describing the proximity effect for any kind of superconducting
bilayer, in terms of thickness, critical temperature and cleanliness, and
superconducting trilayers with a thin third layer has been presented. It has
been proposed that any existing samples, even if very clean, would obey to
these equations which are valid in the dirty limit, because of the presence
of imperfections at the boundary. Only in very specific cases of extremely
thin and smoothly deposited layers (with an rms roughness at the interface
of the order of the interatomic distance in both materials) would the model
not apply. This model has been presented and examined experimentally for
typical values of the parameters involved for tunnel junctions, using
thicknesses intermediate between the extreme cases discussed in previous
publications. Finally, the model has been shown to be very effective in
determining the various important parameters for practical cases, using the
current-voltage characteristics of STJ's (energy gap and critical current as
a function of temperature).

\pagebreak

\section{Tables}

\frame{%
\begin{tabular}{ccccccc}
& RRR$_{b}$ & l$_{0,300}$ & b$_{t}$ & N$\nu $ & T$_{c,0}$ & $\xi _{0}$ \\ 
& / & nm & nm & / & K & nm \\ 
Nb & 25 & 2.2 & 0.26 & 0.35 & 9.2 & 40 \\ 
Al & 4 & 10.3 & 0.2 & 0.175 & 1.2 & 1600
\end{tabular}
}

Table 1: Basic parameters for Nb and Al

\bigskip

\frame{%
\begin{tabular}{ccccc}
& $\gamma $ (th.) & $\gamma $ (exp.) & $\gamma _{BN}$ (th.) & $\gamma _{BN}$
(exp.) \\ 
15nm Al & 1.07 & 0.9 & 1.85 & 1.5 \\ 
30nm Al & 1.32 & 1.2 & 2.3 & 2.2 \\ 
45nm Al & 1.6 & 1.6 & 2.7 & 2.7 \\ 
90nm Al & 1.6 & 1.3 & 2.7 & 2.7
\end{tabular}
}

Table 2: Comparison of theoretical and experimental values of the interface
parameters for Nb/Al STJs with different Al film thicknesses.

\pagebreak

\pagebreak

{\large {\bf Captions for figures}}

{\bf Figure~1:}

Typical output for Al-Nb. Pair potential (a) and density of states (b) for $%
\gamma =1.3$ and $\gamma _{BN}=2.7$, with $d_{S_{2}}=1.7$ $\xi _{2}^{\ast }$
and $d_{S_{1}}=4.35$ $\xi _{1}.$ (a) The upper dashed line is the bulk
energy gap of Nb. The lower dashed line is the bulk energy gap of Al. The
intermediate dashed line is the resulting energy gap, as determined from Fig
1 (b). (b) The densities of states are represented for both materials at the
free interface (solid lines) and at the $S_{1}$-$S_{2}$ interface (dashed
lines).

{\bf Figure~2:}

Typical output for Al-Ta. Pair potential (a) and density of states (b) for $%
\gamma =0.05$ and $\gamma _{BN}=3.$, with $d_{S_{2}}=\xi _{2}^{\ast }$ and $%
d_{S_{1}}=1.18$ $\xi _{1}$. (a) The upper dashed line is the bulk energy gap
of Ta. The lower dashed line is the bulk energy gap of Al. The intermediate
dashed line is the resulting energy gap, as determined from Fig 2 (b). (b)
The densities of states are represented for both materials at the free
interface (solid lines) and at the $S_{1}$-$S_{2}$ interface (dashed lines).

{\bf Figure~3:}

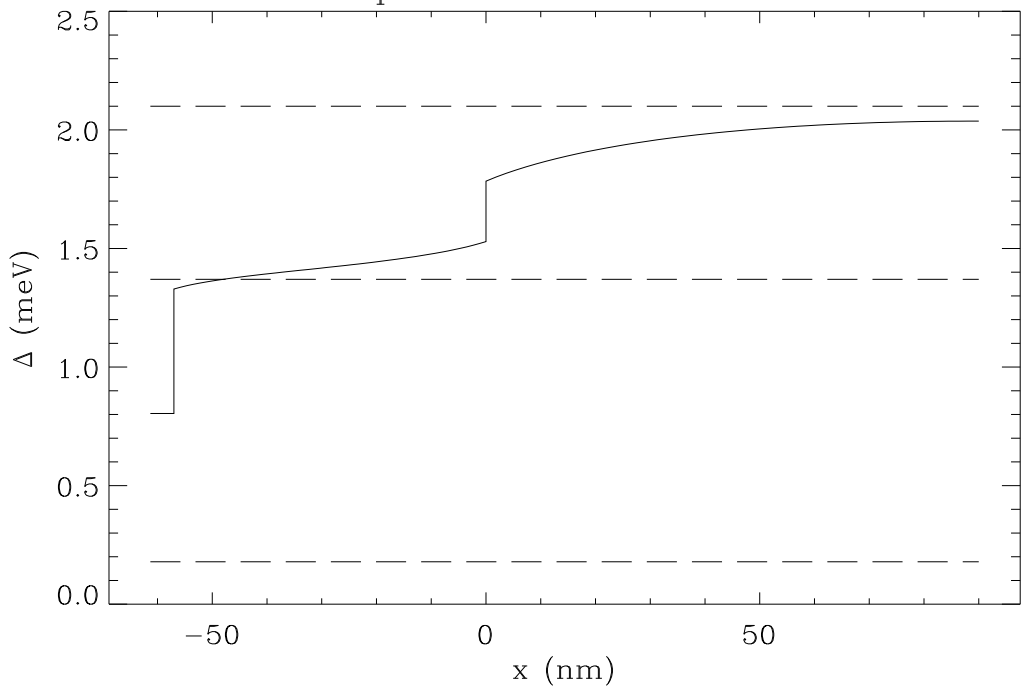
Typical output for Al-Nb-NbN. Pair potential (a) and density of states (b)
for $\gamma _{1}=2.$, $\gamma _{BN_{1}}=1.$, $\gamma _{2}=1.3$ and $\gamma
_{BN_{2}}=2.7$ with $d_{S_{3}}=0.1$ $\xi _{3}^{\ast },$ $d_{S_{2}}=3$ $\xi
_{2}^{\ast }$ and $d_{S_{1}}=3$ $\xi _{1}.$ (a) The upper dashed line is the
bulk energy gap of NbN. The lower dashed line is the bulk energy gap of Al.
The intermediate dashed line is the resulting energy gap, as determined from
Fig 3 (b). (b) The densities of states are represented for NbN at the free
interface (solid line), at the $S_{1}$-$S_{2}$ interface (dashed lines)and
at the $S_{2}-S_{3}$ interface (dotted lines).

{\bf Figure~4:}

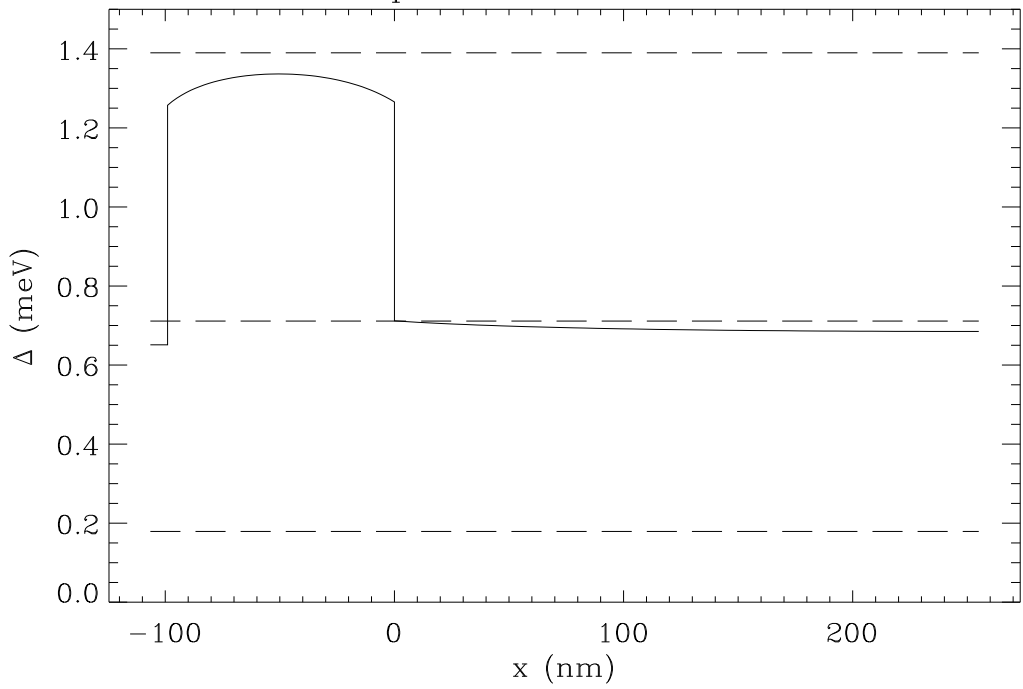
Typical output for Al-Nb-Ta. Pair potential (a) and density of states (b)
for $\gamma _{1}=0.3$, $\gamma _{BN_{1}}=3.8$, $\gamma _{2}=1.3$ and $\gamma
_{BN_{2}}=2.7$ with $d_{S_{3}}=0.1$ $\xi _{3}^{\ast },$ $d_{S_{2}}=0.5$ $\xi
_{2}^{\ast }$ and $d_{S_{1}}=3$ $\xi _{1}.$ (a) The upper dashed line is the
bulk energy gap of Nb. The lower dashed line is the bulk energy gap of Al.
The intermediate dashed line is the resulting energy gap, as determined from
Fig 4 (b). (b) The densities of states are represented for Ta at the free
interface (solid line), at the $S_{1}$-$S_{2}$ interface (dashed lines) and
at the $S_{2}-S_{3}$ interface (dotted lines).

{\bf Figure~5:}

Energy gap (a) and critical current (b) as a function of temperature for the
Al-Nb STJ ($d_{Al}=90nm,$ $d_{Nb}=100nm)$. The data (diamonds) is fitted
using $\gamma =1.3$ and $\gamma _{BN}=2.7.$The expected experimental
uncertainty for the data points is also represented.

{\bf Figure~6:}

Energy gap (a) and critical current (b) as a function of temperature for the
Al-Ta STJ ($d_{Al}=55nm,$ $d_{Ta}=100nm)$. The data (diamonds) is fitted
using $\gamma =0.05$ and $\gamma _{BN}=3.$


\begin{thebibliography}{99}
\bibitem{mcm}  W.L.~McMillan, Phys.Rev. {\bf 175}, 537 (1968).

\bibitem{gol-94}  A.A.~Golubov, E.P.~Houwman, J.G.~Gijsbertsen, J.~Flokstra,
and H.~Rogalla, Phys.Rev.B {\bf 49}, 12953 (1994).

\bibitem{GH-95}  A.A.~Golubov, E.P.~Houwman, J.G.~Gijsbertsen, V.M.~Krasnov,
M.Yu.~Kupriyanov, J.~Flokstra, and H.~Rogalla, Phys.Rev.B {\bf 51}, 1073
(1995).

\bibitem{KL}  M.Yu.~Kupriyanov and V.F.~Lukichev, Zh.Eksp.Teor.Fiz. {\bf \ 94%
}, 139 (1988) [Sov.Phys.JETP {\bf 67}, 1163 (1988)].

\bibitem{GK-89}  A.A.~Golubov and M.Yu.~Kupriyanov, Zh.Eksp.Teor.Fiz. {\bf \
96}, 120 (1989).

\bibitem{Bruder}  W.~Belzig, C.~Bruder and G.~Schon, Phys.Rev.B {\bf \ 54},
9443 (1996).

\bibitem{AGD}  A.A.~Abrikosov, L.P.~Gor'kov, I.E.~Dzyaloshinskii, Methods of
Quantum Field Theory in Statistical Physics, (1962).

\bibitem{Eilenb}  G.~Eilenberger, Z.Phys.{\bf \ 214}, 195 (1968).

\bibitem{Usadel}  K.D.~Usadel, Phys. Rev. Lett. {\bf 25}, 507 (1970).

\bibitem{Eliash}  G.M.~Eliashberg, Zh.Eksp.Teor.Fiz. {\bf 61}, 1254 (1971)
[Sov.Phys.JETP {\bf 34}, 668 (1972)].

\bibitem{LO}  A.I.~Larkin and Yu.N.~Ovchinnikov, Zh.Eksp.Teor.Fiz. {\bf 73},
299 (1977) [Sov.Phys.JETP {\bf 41}, 960 (1977)].

\bibitem{anderson}  P.G.~de~Gennes. Superconductivity of Metals and Alloys
(New York, Benjamin, 1966), p.157.

\bibitem{Aminov}  B. A. Aminov, A.A.~Golubov and M.Yu.~Kupriyanov,
Phys.Rev.B {\bf 53}, 365 (1996).

\bibitem{NbN}  N. Rando, P.Verhoeve, A.Poelaert, A.Peacock, D.J. Goldie, J.
Appl. Phys. {\bf 83}, 10 (1998).

\bibitem{rando92}  N.~Rando, A.~Peacock, A.~van Dordrecht, C.~Foden,
R.~Engelhardt, B.G.~Taylor, P.~Gar\'{e}, J.M.~Lumley, and C.~Pereira,
Nucl.Instr.Met.Ph.Res.A {\bf 313}, 173 (1992).

\bibitem{ver96}  P.~Verhoeve, N.~Rando, J.~Verveer, A.~Peacock,
A.~van~Dordrecht, P.~Videler, M.~Bavdaz, D.J.~Goldie, T.~Lederer,
F.~Scholze, G.~Ulm, and R.~Venn, Phys.Rev.B {\bf 53}, 809 (1996).

\bibitem{mears}  C.A.~Mears, S.E.~Labov, and A.T.~Barfknecht, Low~Temp.Phys. 
{\bf 92}, 561 (1993).

\bibitem{peacock96}  A.~Peacock, P.~Verhoeve, N.~Rando, A.~van~Dordrecht,
B.G.~Taylor, C.~Erd, M.A.C.~Perryman, R.~Venn, J.~Howlett, D.J.~Goldie,
J.~Lumley, and M.~Wallis, Nature {\bf 381}, 135 (1996).

\bibitem{peacock97}  A.~Peacock, P.~Verhoeve, N.~Rando, C.~Erd, M.~Bavdaz,
B.G.~Taylor, and D.~Perez, Astron. Astrophys. Suppl. Ser. {\bf 127}, 497
(1998).

\bibitem{ver97}  P.~Verhoeve, N. Rando, A. Peacock, A. van Dordrecht,
A.~Poelaert, D.J. Goldie, R. Venn, J.Appl.Phys. {\bf 83}, 11, 6118 (1998).

\bibitem{hettl}  P.~Hettl et al., Proceedings of EDXRS-98, Bologna, Italy,
June 7-12, 1(1998).

\bibitem{goldie}  D.~J.~Goldie, P.~L.~Brink, C.~Patel, N.~E.~Booth, and
G.~L.~Salmon, Appl. Phys. Lett. {\bf 64}, 3169 (1994).

\bibitem{dekorte92}  P.~de~Korte, M.L.~van~den~Berg, M.P.~Bruijn,
M.~Frericks, J.B.~le~Grand, J.G.~Gijsbertsen, E.P.~Houwman, and J.~Flokstra,
Proc. SPIE {\bf 1743}, 24 (1992).

\bibitem{kaplan}  S.B.~Kaplan, C.C.~Chi, D.~Langenberg, J.J.~Chang,
S.~Jafarey, and D.~Scalapino, Phys. Rev. B {\bf 14}, 4854 (1976).

\bibitem{booth}  N.~Booth, Appl. Phys. Lett. {\bf 50}, 293 (1987).

\bibitem{poelaert96}  A.~Poelaert, P.~Verhoeve, N.~Rando, and A.~Peacock,
Proc. SPIE {\bf 2808}, 523 (1996).

\bibitem{gray}  K.E.~Gray, J.Phys.F: Metal Phys. {\bf 1}, 290 (1971).

\bibitem{Cooper}  L.N. Cooper,Phys. Rev. Lett. {\bf 6,} 689 (1961).

\bibitem{Movshovitz}  D. Movshovitz, N.Wiser, Phys. Rev. B {\bf 41}, 10503
(1990).

\bibitem{Meservey}  R. Meservey, B. B. Schwartz, , {\it Superconductivity }%
(Vol.1, R.D. Parks (ed.), Dekker, New York,1969), p.12

\bibitem{zehnder}  A. Zehnder, Ph. Lerch, S.P. Zhao, Th. Nussbaumer, E.C.
Kirk, H.R. Ott, Phys. Rev. B, 59 (1999) 8875.
\end{thebibliography}
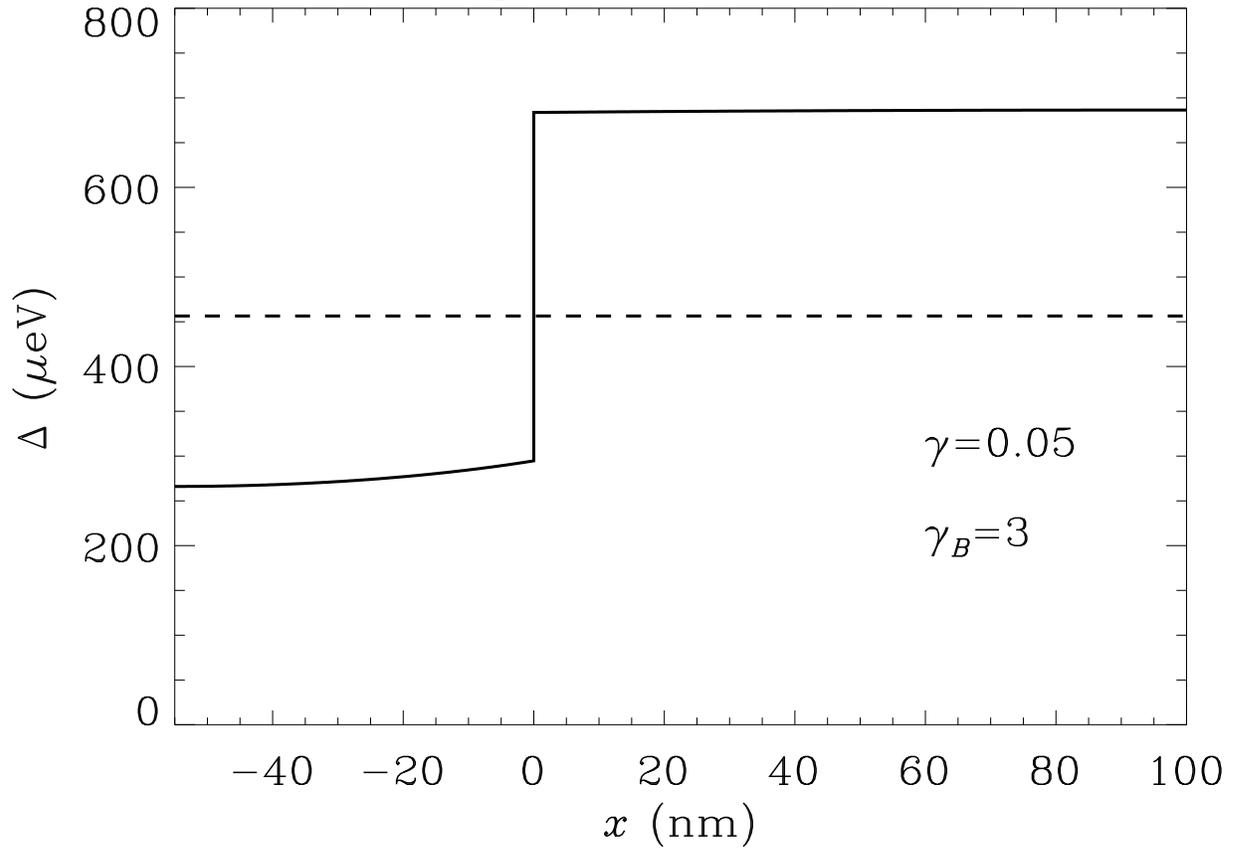
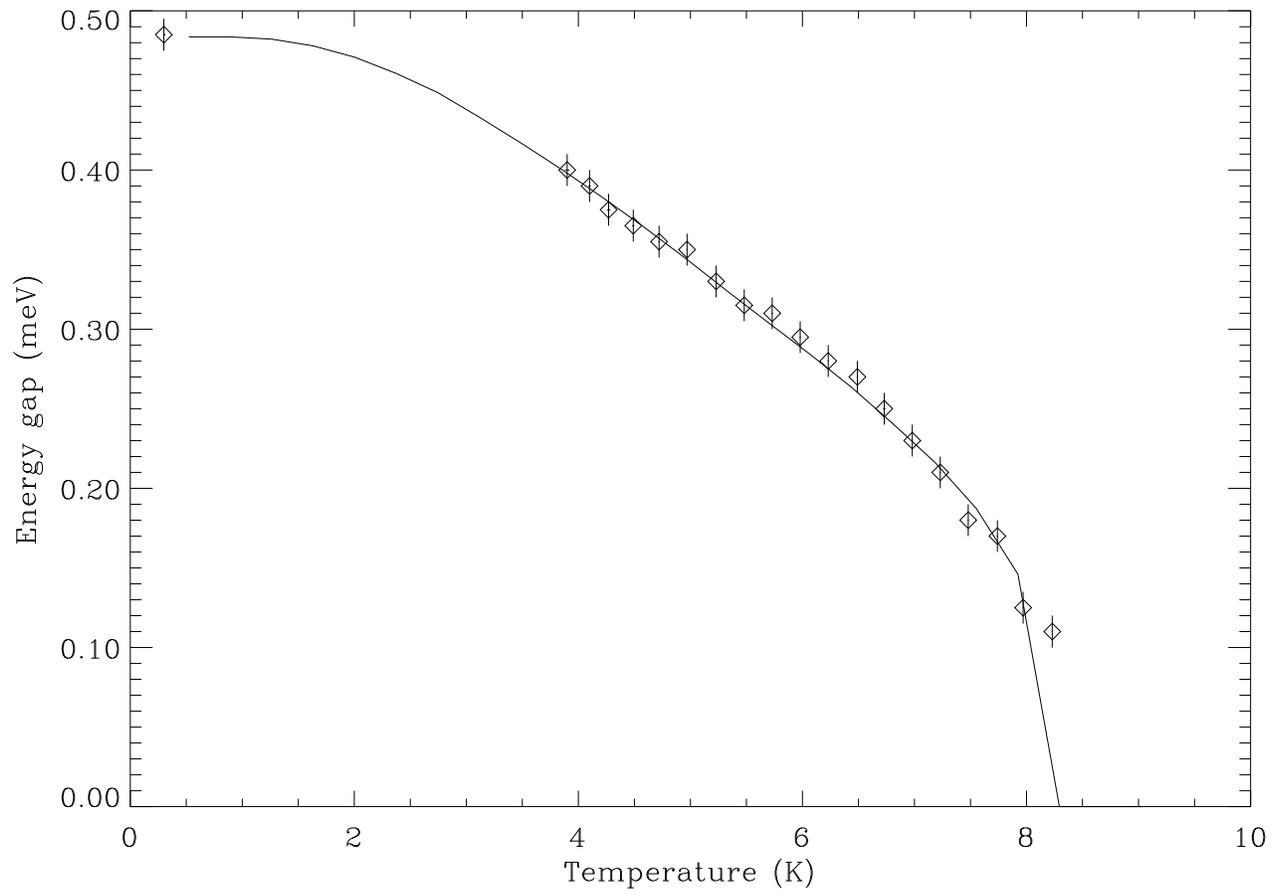
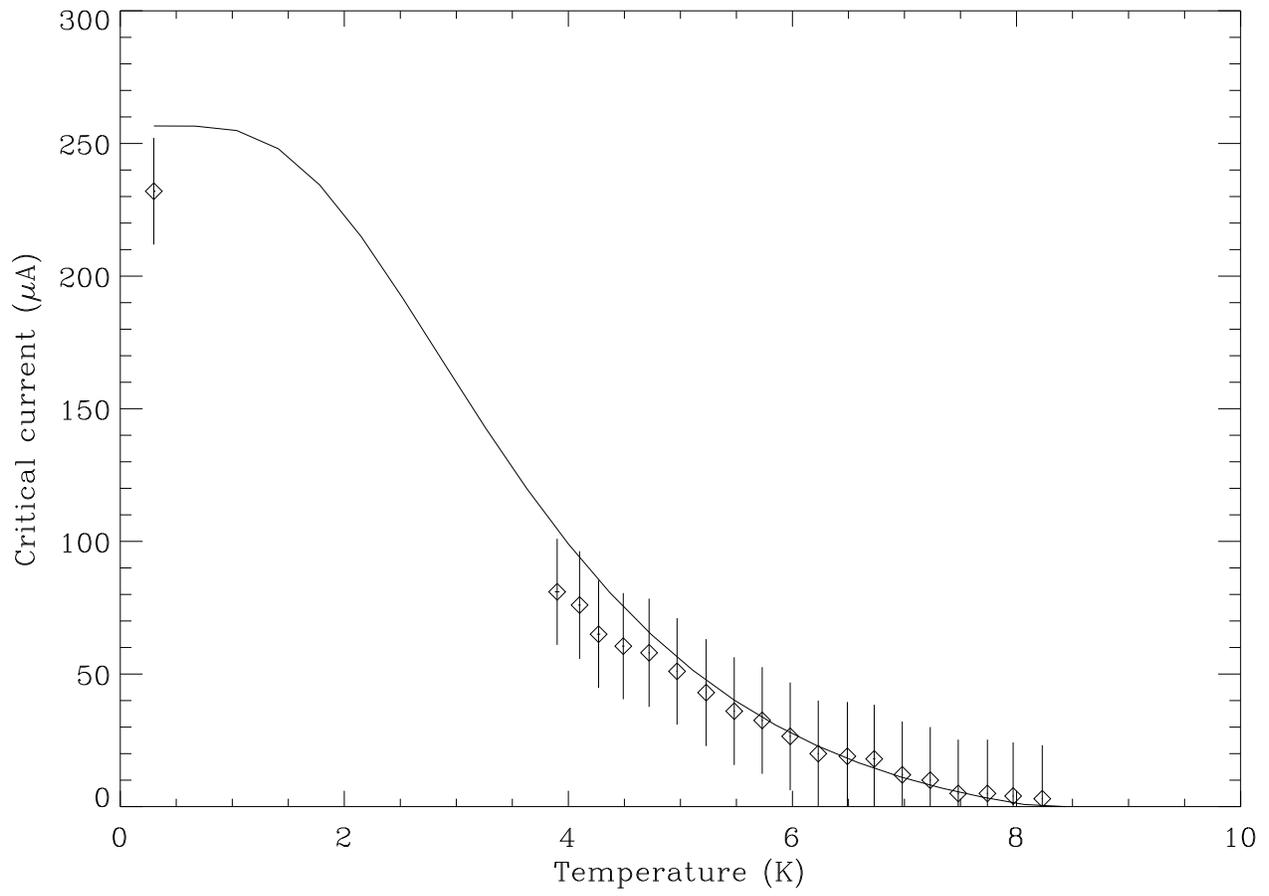
\end{document}